\documentclass{llncs}

\usepackage[english]{babel}
\usepackage[T1]{fontenc}
\usepackage{lmodern}
\usepackage[utf8]{inputenc}
\usepackage[final]{microtype}

\usepackage{amsmath,amssymb,xcolor,graphicx,wrapfig,paralist}
\usepackage{algorithmicx,algorithm}
\usepackage[noend]{algpseudocode}
\usepackage{booktabs}
\usepackage{proof}
\usepackage{tikz}
\usepackage{csquotes}
\usepackage{graphics}
\usepackage{stmaryrd}
\usepackage[caption=false]{subfig}

\usepackage{ellipsis, mparhack, ragged2e} 
\usepackage[l2tabu, orthodox]{nag} 

\usetikzlibrary{shapes,positioning,arrows,calc,automata,matrix,fit}
\tikzstyle{state}+=[minimum size = 6mm, inner sep=0,outer sep=1]
\tikzset{->,>=stealth'}

\usepackage{tabu,multirow}
\newcolumntype{L}{X}
\newcolumntype{R}{>{\raggedleft\arraybackslash}X}
\newcolumntype{C}{>{\centering\arraybackslash}X}

\input{macros}

\newcommand{\todojan}[1]{\todo{\textbf{J:\\}#1}}
\newcommand{\todoprzemek}[1]{\todo{\textbf{P:\\}#1}}
\newcommand{\todopranav}[1]{\todo{\textbf{Pranav:\\}#1}}
\newcommand{\todotobias}[1]{\todo{\textbf{Tobias:\\}#1}}

\renewcommand{\todojan}[1]{}
\renewcommand{\todoprzemek}[1]{}
\renewcommand{\todopranav}[1]{}
\renewcommand{\todotobias}[1]{}

\newcommand{\rewformat}[1]{\textbf{#1}}

\title{Value Iteration for Long-run Average Reward in Markov Decision Processes}

\author{Pranav Ashok\inst1 \and
	Krishnendu Chatterjee\inst2 \and
	Przemys{\l}aw Daca\inst2 \and
	Jan K{\v{r}}et\'insk\'y\inst1 \and
	Tobias Meggendorfer\inst1} %
\institute{Technical University of Munich, Germany \and
	IST Austria} %

\begin{document}

\maketitle

\begin{abstract}
	Markov decision processes (MDPs) are standard models for probabilistic systems with non-deterministic behaviours.
	Long-run average rewards provide a mathematically elegant formalism for expressing long term performance.
	Value iteration (VI) is one of the simplest and most efficient algorithmic approaches to MDPs with other properties, such as reachability objectives.
	Unfortunately, a naive extension of VI does not work for MDPs with long-run average rewards, as there is no known stopping criterion.
	In this work our contributions are threefold.
	(1)~We refute a conjecture related to stopping criteria for MDPs with long-run average rewards.
	(2)~We present two practical algorithms for MDPs with long-run average rewards based on VI.
	First, we show that a combination of applying VI locally for each maximal end-component (MEC) and VI for reachability objectives can provide approximation guarantees.
	Second, extending the above approach with a simulation-guided on-demand variant of VI, we present an anytime algorithm that is able to deal with very large models.
	(3)~Finally, we present experimental results showing that our methods significantly outperform the standard approaches on several benchmarks.
\end{abstract}

\section{Introduction}

The analysis of probabilistic systems arises in diverse application contexts of computer science, e.g. analysis of randomized communication and security protocols, stochastic distributed systems, biological systems, and robot planning, to name a few.
The standard model for the analysis of probabilistic systems that exhibit both probabilistic and non-deterministic behaviour are \emph{Markov decision processes (MDPs)}~\cite{Howard,FV97,Puterman}.
An MDP consists of a finite set of states, a finite set of actions, representing the non-deterministic choices, and a transition function that given a state and an action gives the probability distribution over the successor states.
In verification, MDPs are used as models for e.g.\ concurrent probabilistic systems~\cite{CY95} or probabilistic systems operating in open environments~\cite{SegalaT}, and are applied in a wide range of applications~\cite{BaierBook,KNP11}.

\paragraph{Long-run average reward}
A \emph{payoff} function in an MDP maps every infinite path (infinite sequence of state-action pairs) to a real value.
One of the most well-studied and mathematically elegant payoff functions is the \emph{long-run average reward} (also known as \emph{mean-payoff} or \emph{limit-average reward}, \emph{steady-state reward} or simply \emph{average reward}), where every state-action pair is assigned a real-valued reward, and the payoff of an infinite path is the long-run average of the rewards on the path~\cite{FV97,Puterman}.
Beyond the elegance, the long-run average reward is standard to model performance properties, such as the average delay between requests and corresponding grants, average rate of a particular event, etc.
Therefore, determining the maximal or minimal expected long-run average reward of an MDP is a basic and fundamental problem in the quantitative analysis of probabilistic systems.

\paragraph{Classical algorithms}
A \emph{strategy} (also known as \emph{policy} or \emph{scheduler}) in an MDP specifies how the non-deterministic choices of actions are resolved in every state.
The \emph{value} at a state is the maximal expected payoff that can be guaranteed among all strategies.
The values of states in MDPs with payoff defined as the long-run average reward can be computed in polynomial-time using linear programming~\cite{FV97,Puterman}.
The corresponding linear program is quite involved though.
The number of variables is proportional to the number of state-action pairs and the overall size of the program is linear in the number of transitions (hence potentially quadratic in the number of actions).
While the linear programming approach gives a polynomial-time solution, it is quite slow in practice and does not scale to larger MDPs.
Besides linear programming, other techniques are considered for MDPs, such as dynamic-programming through strategy iteration or value iteration~\cite[Chap.~9]{Puterman}.

\paragraph{Value iteration}
A generic approach that works very well in practice for MDPs with other payoff functions is \emph{value iteration (VI)}.
Intuitively, a particular one-step operator is applied iteratively and the crux is to show that this iterative computation converges to the correct solution (i.e.\ the value).
The key advantages of VI are the following:
\begin{enumerate}
	\item \emph{Simplicity.}
	VI provides a very simple and intuitive dynamic-programming algorithm which is easy to adapt and extend.
	\item \emph{Efficiency.}
	For several other payoff functions, such as finite-horizon rewards (instantaneous or cumulative reward) or reachability objectives, applying the concept of VI yields a very efficient solution method.
	In fact, in most well-known tools such as PRISM~\cite{KNP11}, value iteration performs much better than linear programming methods for reachability objectives.
	\item \emph{Scalability.}
	The simplicity and flexibility of VI allows for several improvements and adaptations of the idea, further increasing its performance and enabling quick processing of very large MDPs.
	For example, when considering reachability objectives, \cite{DBLP:conf/ijcai/PineauGT03} present point-based value-iteration (PBVI), applying the iteration operator only to a part of the state space, and~\cite{DBLP:conf/icml/McMahanLG05} introduce bounded real-time dynamic programming (BRTDP), where again only a fraction of the state space is explored based on partial strategies.
	Both of these approaches are simulation-guided, where simulations are used to decide how to explore the state space.
	The difference is that the former follows an offline computation, while the latter is online.
	Both scale well to large MDPs and use VI as the basic idea to build upon.
\end{enumerate}

\paragraph{Value iteration for long-run average reward}
While VI is standard for reachability objectives or finite-horizon rewards, it does not work for general MDPs with long-run average reward.
The two key problems pointed out in~\cite[Sect.~8.5,~9.4]{Puterman} are as follows:
(a)~if the MDP has some periodicity property, then VI does not converge;
and (b)~for general MDPs there are neither bounds on the speed of convergence nor stopping criteria to determine when the iteration can be stopped to guarantee approximation of the value.
The first problem can be handled by adding self-loop transitions~\cite[Sect.~8.5.4]{Puterman}.
However, the second problem is conceptually more challenging, and a solution is conjectured in~\cite[Sect.~9.4.2]{Puterman}.

\paragraph{Our contribution}
In this work, our contributions are related to value iteration for MDPs with long-run average reward, they range from conceptual clarification to practical algorithms and experimental results.
The details of our contributions are as follows.
\begin{itemize}
	\item \emph{Conceptual clarification.}
	We first present an example to refute the conjecture of~\cite[Sect.~9.4.2]{Puterman}, showing that the approach proposed there does not suffice for VI on MDPs with long-run average reward.
	\item \emph{Practical approaches.}
	We develop, in two steps, practical algorithms instantiating VI for approximating values in MDPs with long-run average reward.
	Our algorithms take advantage of the notion of maximal end-components (MECs) in MDPs. 
	Intuitively, MECs for MDPs are conceptually similar to strongly connected components (SCCs) for graphs and recurrent classes for Markov chains.
	We exploit these MECs to arrive at our two methods:
	\begin{enumerate}
		\item The first variant applies VI locally to each MEC in order to obtain an approximation of the values within the MEC.
		After the approximation in every MEC, we apply VI to solve a reachability problem in a modified MDP with collapsed MECs.
		We show that this simple combination of VI approaches ensures guarantees on the approximation of the value.
		\item We then build on the approach above  to present a simulation-guided variant of VI.
		In this case, the approximation of values for each MEC and the reachability objectives are done at the same time using VI.
		For the reachability objective a BRDTP-style VI (similar to~\cite{atva}) is applied, and within MECs VI is applied on-demand (i.e. only when there is a requirement for more precise value bounds).
		The resulting algorithm furthermore is an \emph{anytime} algorithm, i.e.\ it can be stopped at any time and give an upper and lower bounds on the result.
	\end{enumerate}
	\item \emph{Experimental results.}
	We compare our new algorithms to the state-of-the-art tool MultiGain~\cite{DBLP:conf/tacas/BrazdilCFK15} on various models.
	The experiments show that MultiGain is vastly outperformed by our methods on nearly every model.
	Furthermore, we compare several variants of our methods and investigate the different domains of applicability.
\end{itemize}
In summary, we present the first instantiation of VI for general MDPs with long-run average reward.
Moreover, we extend it with a simulation-based approach to obtain an efficient algorithm for large MDPs.
Finally, we present experimental results demonstrating that these methods provide significant improvements over existing ones.

\paragraph{Further related work.}
There is a number of techniques to compute or approximate the long-run average reward in MDPs~\cite{Puterman,Howard,veinott1966}, ranging from linear programming to value iteration to strategy iteration.
Symbolic and explicit techniques based on strategy iteration are combined in~\cite{DBLP:conf/qest/WimmerBBHCHDT10}.
Further, the more general problem of MDPs with multiple long-run average rewards was first considered in~\cite{Cha07}, a complete picture was presented in~\cite{krish,DBLP:conf/lics/ChatterjeeKK15} and partially implemented in~\cite{DBLP:conf/tacas/BrazdilCFK15}.
The extension of our approach to multiple long-run average rewards, or combination of expectation and variance~\cite{BCFK13}, are interesting directions for future work.
Finally, VI for MDPs with guarantees for reachability objectives was considered in~\cite{atva,haddad2014reachability}.

Proofs and supplementary material can be found in \cite{techrep}. 

%
%

\section{Preliminaries} \label{sec:prelim}

\subsection{Markov decision processes}
A \emph{probability distribution} on a finite set $X$ is a mapping $\rho: X\mapsto [0,1]$, such that $\sum_{x\in X} \rho(x) = 1$. We denote by $\Distributions(X)$ the set of all probability distributions on $X$.
Further, the \emph{support} of a probability distribution $\rho$ is denoted by $\supp(\rho) = \{x \in X \mid \rho(x)> 0 \}$.
\begin{definition}[MDP]
	A \emph{Markov decision processes (MDP)} is a tuple of the form $\Mdp = (\states, \initstate, \actions, \av, \trans, \rew)$, where $\states$ is a finite set of \emph{states}, $\initstate \in \states$ is the \emph{initial} state, $\actions$ is a finite set of \emph{actions}, $\av: \states \to 2^{\actions}$ assigns to every state a set of \emph{available} actions, $\trans: \states \times \actions \to \distributions(\states)$ is a \emph{transition function} that given a state $s$ and an action $a\in \av(s)$ yields a probability distribution over successor states, and $\rew : \states \times \actions \to \Reals^{\geq 0}$ is a \emph{reward function}, assigning rewards to state-action pairs.
\end{definition}
%
For ease of notation, we write $\trans(s, a, s')$ instead of $\trans(s, a)(s')$.

An \emph{infinite path} $\path$ in an MDP is an infinite word $\path = s_0 a_0 s_1 a_1 \dots \in (\states \times \actions)^\omega$, such that for every $i \in \Naturals$, $a_i\in \av(s_i)$ and $\trans(s_i,a_i, s_{i+1}) > 0$.
A \emph{finite path} $\fpath = s_0 a_0 s_1 a_1 \dots s_n \in (\states \times \actions)^* \times \states$ is a finite prefix of an
infinite path.

A \emph{strategy} on an MDP is a function $\straa: (\states \times \actions)^*\times S \to \distributions(\actions)$, which given a finite path $\fpath = s_0 a_0 s_1 a_1 \dots s_n$ yields a probability distribution $\straa(\fpath) \in \distributions(\av(s_n))$ on the actions to be taken next.
We call a strategy \emph{memoryless randomized} (or \emph{stationary}) if it is of the form $\straa: \states \to \distributions(\actions)$, and \emph{memoryless deterministic} (or \emph{positional}) if it is of the form $\straa: \states \to \actions$.
We denote the set of all strategies of an MDP by $\straas$, and the set of all memoryless deterministic strategies by $\straas^{\mathsf{MD}}$.
Fixing a strategy $\straa$ and an initial state $s$ on an MDP $\Mdp$ gives a unique probability measure $\pr^\straa_{\Mdp, s}$ over infinite paths~\cite[Sect.~2.1.6]{Puterman}.
The expected value of a random variable $F$ is defined as $\expected^\straa_{\Mdp, s}[F] = \int F\ d\,\pr^\straa_{\Mdp, s}$.
When the MDP is clear from the context, we drop the corresponding subscript and write $\pr^\straa_s$ and $\expected^\straa_s$ instead of $\pr^\straa_{\Mdp, s}$ and $\expected^\straa_{\Mdp, s}$, respectively.

\paragraph{End components}
A pair $(T, A)$, where $\emptyset \neq T \subseteq S$ and $\emptyset \neq A \subseteq \Union_{s\in T} \av(s)$, is an \emph{end component} of an MDP $\Mdp$ if (i)~for all $s \in T, a \in A \intersection \av(s)$ we have $\supp(\trans(s,a)) \subseteq T$, and (ii)~for all $s, s' \in T$ there is a finite path $\fpath = s a_0 \dots a_n s' \in (T \times A)^* \times T$, i.e.\ $\fpath$ starts in $s$, ends in $s'$, stays inside $T$ and only uses actions in $A$.\footnote{This standard definition assumes that actions are unique for each state, i.e.\ $\av(s) \intersection \av(s') = \emptyset$ for $s \neq s'$.
The usual procedure of achieving this in general is to replace $\actions$ by $\states \times \actions$ and adapting $\av$, $\trans$, and $\rew$ appropriately.}
Intuitively, an end component describes a set of states for which a particular strategy exists such that all possible paths remain inside these states and all of those states are visited infinitely often almost surely.
An end component $(T, A)$ is a \emph{maximal end component (MEC)} if there is no other end component $(T', A')$ such that $T \subseteq T'$ and $A \subseteq A'$.
Given an MDP $\Mdp$, the set of its MECs is denoted by $\mec(\Mdp)$.
With these definitions, every state of an MDP belongs to at most one MEC and each MDP has at least one MEC.

Using the concept of MECs, we recall the standard notion of a \emph{MEC quotient}~\cite{DeAlfaro1997}.
To obtain this quotient, all MECs are merged into a single representative state, while transitions between MECs are preserved.
Intuitively, this abstracts the MDP to its essential infinite time behaviour.

\begin{definition}[MEC quotient~\cite{DeAlfaro1997}] \label{def:mq}
	Let $\Mdp = (\states, \initstate, \actions, \av, \trans, \rew)$ be an MDP with MECs $\mec(\Mdp) = \{(T_1, A_1), \dots, (T_n, A_n)\}$.
	Further, define $\mec_\states = \Union_{i=1}^n T_i$ as the set of all states contained in some MEC.
	The \emph{MEC quotient of} $\Mdp$ is defined as the MDP $\widehat\Mdp = (\widehat\states, \widehat{s}_\textrm{init}, \widehat\actions, \widehat\av, \widehat\trans, \widehat\rew)$, where:
	\begin{itemize}
		\item $\widehat\states = \states \setminus \mec_\states \union \set{\widehat{s}_1, \dots, \widehat{s}_n}$,
		\item if for some $T_i$ we have $\initstate\in T_i$, then $\widehat{s}_\textrm{init} = \widehat{s}_i$,  otherwise $\widehat{s}_\textrm{init} = \initstate$,
		\item $\widehat\actions =  \set{(s,a) \mid s\in \states, a\in \av(s)}$,
		\item the available actions $\widehat\av$ are defined as
		\begin{align*}
			\forall s \in \states \setminus \mec_\states.~& \widehat\av(s) = \set{(s,a) \mid a \in \av(s)} \\
			\forall 1 \leq i \leq n.~& \widehat\av(\widehat{s}_i) = \set{(s,a) \mid s \in T_i \land a \in \av(s) \setminus A_i},
		\end{align*}
		\item the transition function $\widehat\trans$ is defined as follows.
		Let $\widehat{s} \in \widehat{S}$ be some state in the quotient and $(s, a) \in \av(\widehat{s})$ an action available in $\widehat{s}$.
		Then
		\begin{equation*}
			\widehat{\trans}(\widehat{s}, (s, a), \widehat{s}') = \begin{dcases*}
				{\sum}_{s' \in T_j} \trans(s, a, s') & if $\widehat{s}' = \widehat{s}_j$, \\
				\trans(s, a, \widehat{s}') & otherwise, i.e.\ $\widehat{s}' \in \states \setminus \mec_\states$.
			\end{dcases*}
		\end{equation*}
		For the sake of readability, we omit the added self-loop transitions of the form $\trans(\widehat{s}_i, (s, a), \widehat{s}_i)$ with $s \in T_i$ and $a \in A_i$ from all figures.
		\item Finally, for $\widehat{s} \in \widehat{\states}$, $(s, a) \in \widehat{\av}(\widehat{s})$, we define $\widehat\rew(s, (s, a)) = \rew(s, a)$.
	\end{itemize}
	Furthermore, we refer to $\widehat{s}_1, \dots, \widehat{s}_n$ as \emph{collapsed states} and identify them with the corresponding MECs.
\end{definition}
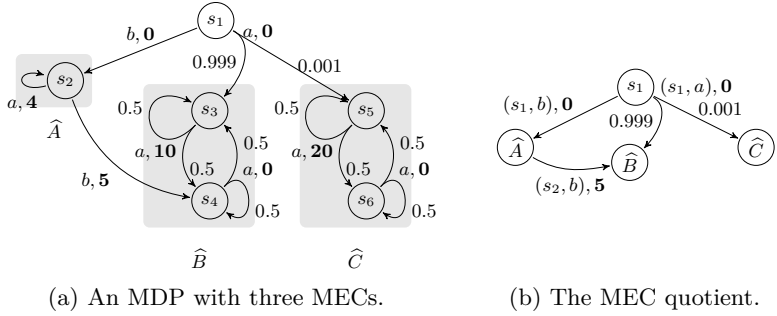
\begin{figure}[t]
	\centering
	\subfloat[An MDP with three MECs.]{\label{fig:example}
		\scalebox{0.8}{
			\begin{tikzpicture}[auto]
				\node[state] (s1) at (-3,0.5) {$s_1$};
				\node[state] (s6) at (-0.5,-2.5) {$s_6$};
				\node[state] (s5) at (-0.5,-1) {$s_5$};
				\node[state] (s3) at (-3.1,-1) {$s_3$};
				\node[state] (s4) at (-3.1,-2.5) {$s_4$};
				\node[state] (s2) at (-5.5,-0.5) {$s_2$};
				\node[] at (-2.3,0.3) {$a, \rewformat{0}$};
				\node[] at ($(s4) + (0.8,0.5)$) {$a,\rewformat{0}$};
				\node[] at ($(s6) + (0.8,0.5)$) {$a, \rewformat{0}$};
				\node[] at ($(s3) + (-0.9,-0.7)$) {$a, \rewformat{10}$};
				\node[] at ($(s5) + (-0.9,-0.7)$) {$a, \rewformat{20}$};
			
				\path[->]
					(s1) edge node[right]{$0.001$} (s5)
					(s1) edge[out=-30,in=45] node[left]{$0.999$} (s3)
					(s1) edge node[above] {$b, \rewformat{0}$} (s2)
					(s5) edge[out=225,in=135] node[below right]{0.5} (s6)
					(s5) edge[out=225,in=155,looseness=8] coordinate (e5loop) node[above left]{$0.5$} (s5)
					(s6) edge[out=45,in=-35,looseness=4] coordinate (e6loop) node[below right] {$0.5$} (s6)
					(s6) edge[out=45,in=-45] node[above right] {$0.5$} (s5)
					(s3) edge[out=225,in=135] node[below right]{0.5} (s4)
					(s3) edge[out=225,in=155,looseness=8] coordinate (e3loop) node[above left]{$0.5$} (s3)
					(s4) edge[out=45,in=-35,looseness=4] coordinate (e4loop) node[below right] {$0.5$} (s4)
					(s4) edge[out=45,in=-45] node[above right] {$0.5$} (s3)
					(s2) edge[loop left] coordinate (e2loop) node[below, inner sep=7pt]{$a, \rewformat{4}$} (s2)
					(s2) edge[out=-80,in=180,bend right] node[below left] {$b, \rewformat{5}$} (s4);
				
				\node[rectangle,rounded corners=3pt,draw=none,fill=black,fill opacity=0.1,fit=(s2) (e2loop)] (rectA) {};
				\node[rectangle,rounded corners=3pt,draw=none,fill=black,fill opacity=0.1,fit=(s3) (s4) (e3loop) (e4loop)] (rectB) {};
				\node[rectangle,rounded corners=3pt,draw=none,fill=black,fill opacity=0.1,fit=(s5) (s6) (e5loop) (e6loop)] (rectC) {};
				\node[] at ($(rectA) + (0,-0.8)$) {$\widehat{A}$};
				\node[] at ($(rectB) + (0,-1.7)$) {$\widehat{B}$};
				\node[] at ($(rectC) + (0,-1.7)$) {$\widehat{C}$};
			\end{tikzpicture}
		}
	}
	\quad
	\subfloat[The MEC quotient.]{\label{fig:quo}
		\scalebox{0.8}{
			\begin{tikzpicture}[auto, align=center]
				\node[state] (s0) at (-3,0.5) {$s_1$};
				\node[state] (s2) at (-1,-0.5) {$\widehat C$};
				\node[state] (s3) at (-3.1,-0.75) {$\widehat B$};
				\node[state] (s5) at (-5,-0.5) {$\widehat A$};
				
				\node[] at (-2,0.5) {$(s_1,a), \rewformat{0}$};
				
				\path[->]
				(s0) edge node[anchor=south west,inner sep=1pt] {$0.001$} (s2)
				(s0) edge[out=-30,in=45] node[left]{$0.999$} (s3)
				(s0) edge node[above left, inner sep=1pt] {$(s_1, b), \rewformat{0}$} (s5)
				(s5) edge[out=-30,in=190] node[below] {$(s_2, b), \rewformat{5}$} (s3);
				
				\node[] at (-2,-2.5) {}; 
			\end{tikzpicture}
		}
	}
	\caption{An example of how the MEC quotient is constructed.
		By $a, \rewformat{r}$ we denote that the action $a$ yields a reward of $\rewformat{r}$.}
\end{figure}
\begin{example}
	Figure~\ref{fig:example} shows an MDP with three MECs, $\widehat{A} = (\set{s_2}, \set{a}), \widehat{B} = (\set{s_3,s_4}, \set{a}), \widehat{C} = (\set{s_5,s_6}, \set{a}))$.
	Its MEC quotient is shown in Figure~\ref{fig:quo}.
	\qee
\end{example}
\begin{remark}
	In general, the MEC quotient does not induce a DAG-structure, since there might be probabilistic transitions between MECs.
	Consider for example the MDP obtained by setting $\trans(s_2, b, s_4) = \set*{s_1 \mapsto \frac{1}{2}, s_2 \mapsto \frac{1}{2}}$ in the MDP of Figure~\ref{fig:example}.
	Its MEC quotient then has $\widehat{\trans}(\widehat{A}, (s_2, b)) = \set*{s_1 \mapsto \frac{1}{2}, \widehat{B} \mapsto \frac{1}{2}}$.
\end{remark}
\begin{remark}\label{rem:endcomp}
	The MEC decomposition of an MDP $\Mdp$, i.e.\ the computation of $\mec(\Mdp)$, can be achieved in polynomial time~\cite{CY95}.
	For improved algorithms on general MDPs and various special cases see~\cite{ChatterjeeH11,ChatterjeeH12,ChatterjeeH14,CL13}.
\end{remark}
\begin{definition}[MEC restricted MDP]
	Let $\Mdp$ be an MDP and $(T, A) \in \mec(\Mdp)$ a MEC of $\Mdp$.
	By picking some initial state $\initstate' \in T$, we obtain the \emph{restricted MDP} $\Mdp' = (T, \initstate', A, \av', \trans', \rew')$ where
	\begin{itemize}
		\item $\av'(s) = \av(s) \intersection A$ for $s \in T$,
		\item $\trans'(s, a, s') = \trans(s, a, s')$ for $s, s' \in T$, $a \in A$, and
		\item $\rew'(s, a) = \rew(s, a)$ for $s \in T$, $a \in A$.
	\end{itemize}
\end{definition}
\paragraph{Classification of MDPs}
If for some MDP $\Mdp$, $(\states,\actions)$ is a MEC, we call the MDP \emph{strongly connected}.
If it contains a single MEC plus potentially some transient states, it is called \emph{(weakly) communicating}.
Otherwise, it is called \emph{multichain}~\cite[Sect.~8.3]{Puterman}.

For a Markov chain, let $\trans^n(s, s')$ denote the probability of going from the state $s$ to state $s'$ in $n$ steps.
The \emph{period $p$ of a pair} $s, s'$ is the greatest common divisor of all $n$'s with $\trans^n(s, s') > 0$.
The pair $s, s'$ is called \emph{periodic} if $p > 1$ and \emph{aperiodic} otherwise.
A Markov chain is called aperiodic if all pairs $s, s'$ are aperiodic, otherwise the chain is called periodic.
Similarly, an MDP is called aperiodic if every memoryless randomized strategy induces an aperiodic Markov chain, otherwise the MDP is called periodic.

\paragraph{Long-run average reward}
In this work, we consider the (maximum) \emph{long-run average reward} (or \emph{mean-payoff}) of an MDP, which intuitively describes the (maximum) average reward per step we expect to see when simulating the MDP for time going to infinity.
Formally, let $R_i$ be a random variable, which for an infinite path $\path = s_0 a_0 s_1 a_1 \dots$ returns $R_i(\path) = \rew(s_i, a_i)$, i.e.\ the reward observed at step $i \geq 0$.
Given a strategy $\straa$, the $n$-step average reward then is
\begin{equation*}
	\gain^\straa_n(s) := \expected^\straa_s \left (\frac1n\sum_{i=0}^{n-1} R_i \right),
\end{equation*}
and the \emph{long-run average reward} of the strategy $\straa$ is
\begin{equation*}
	\gain^\straa(s) := \liminf_{n\to\infty} \gain^\straa_n.
\end{equation*}
The $\liminf$ is used in the definition, since the limit may not exist in general for an arbitrary strategy.
Nevertheless, for finite MDPs the optimal limit-inferior (also called the \emph{value}) is attained by some memoryless deterministic strategy $\straa^* \in \straas^{\mathsf{MD}}$ and is in fact the limit~\cite[Thm.~8.1.2]{Puterman}.
\begin{equation*}
	\gain(s) := \sup_{\straa \in \straas} \liminf_{n\to\infty} \expected^\straa_s \left (\frac1n \sum_{i=0}^{n-1} R_i \right) = \sup_{\straa \in \straas} \gain^\straa(s) = \max_{\straa \in \straas^{\mathsf{MD}}} \gain^\straa(s) = \lim_{n\to\infty} \gain^{\straa^*}_n.
\end{equation*}
An alternative well-known characterization we use in this paper is
\begin{equation}
	\gain(s) = \max_{\straa \in \straas^{\mathsf{MD}}} \sum_{M \in \MECs} \pr^\straa_s[\Diamond\Box M] \cdot \gain(M), \label{eq:decomp}
\end{equation}
where $\Diamond\Box M$ denotes the set of paths that eventually remain forever within $M$ and $\gain(M)$ is the unique value achievable in the MDP restricted to the MEC $M$.
Note that $\gain(M)$ does not depend on the initial state chosen for the restriction.
\section{Value Iteration Solutions} \label{sec:vi}

\subsection{Naive value iteration} \label{sec:nvi}

Value iteration is a dynamic-programming technique applicable in many contexts.
It is based on the idea of repetitively updating an approximation of the value for each state using the previous approximates until the outcome is precise enough.
The standard value iteration for average reward~\cite[Sect.~8.5.1]{Puterman} is shown in Algorithm~\ref{alg:vi}.

\begin{algorithm}[t]
	\caption{\textsc{ValueIteration}}
	\label{alg:vi}
	\begin{algorithmic}[1]
		\Require MDP $\Mdp = (\states,\initstate,\actions,\av,\trans,\rew)$, precision $\varepsilon > 0$
		\Ensure $w$, s.t.  $\abs{w - v(\initstate)} < \varepsilon$
		\State $t_0(\cdot) \gets 0$, $n \gets 0$.
		\While {stopping criterion not met}
		\State $n \gets n+1$
		\For {$s \in \states$}
			\State $t_n(s) = \max_{a\in \av(s)} \left(\rew(s,a) + \sum_{s'\in \states} \trans(s, a, s') t_{n-1}(s')\right)$
		\EndFor
		\EndWhile
		\State \Return $\frac1n t_n(\initstate)$
	\end{algorithmic}
\end{algorithm}
First, the algorithm sets $t_0(s) = 0$ for every $s \in \states$.
Then, in the inner loop, the value $t_n$ is computed from the value of $t_{n-1}$ by choosing the action which maximizes the expected reward plus successor values.
This way, $t_n$ in fact describes the optimal \emph{expected $n$-step total reward}
\begin{equation*}
	t_n(s) = \max_{\straa \in \straas^{\mathsf{MD}}} \expected^\straa_s \left( \sum_{i=0}^{n-1} R_i \right) = n \cdot \max_{\straa \in \straas^{\mathsf{MD}}} \gain^\straa_n(s).
\end{equation*}
Moreover, $t_n$ approximates the $n$-multiple of the long-run average reward
\begin{theorem}[{\cite[Thm.~9.4.1]{Puterman}}] \label{th:vi_conv}
	For any MDP $\Mdp$ and any $s \in \states$ we have $\lim_{n \to \infty} \frac1n t_n(s) = \gain(s)$ for $t_n$ obtained by Algorithm~\ref{alg:vi}.
\end{theorem}
\subsubsection{Stopping criteria}
The convergence property of Theorem~\ref{th:vi_conv} is not enough to make the algorithm practical, since it is not known when to stop the approximation process in general.
For this reason, we discuss stopping criteria which describe when it is safe to do so.
More precisely, for a chosen $\varepsilon > 0$ the stopping criterion guarantees that when it is met, we can provide a value $w$ that is $\varepsilon$-close to the average reward $\gain(\initstate)$.

We recall a stopping criterion for communicating MDPs defined and proven correct in~\cite[Sect.~9.5.3]{Puterman}.
Note that in a communicating MDP, all states have the same average reward, which we simply denote by $\gain$.
For ease of notation, we enumerate the states of the MDP $\states = \set{s_1, \dots, s_n}$ and treat the function $t_n$ as a vector of values $\vec{t}_n = (t_n(s_1), \dots, t_n(s_{n}))$.
Further, we define the relative difference of the value iteration iterates as $\vec{\Delta}_{n} := \vec{t}_{n} - \vec{t}_{n-1}$ and introduce the \emph{span semi-norm}, which is defined as the difference between the maximum and minimum element of a vector $\vec{w}$
\begin{equation*}
	\spannorm{\vec{w}} = \max_{s \in \states} \vec{w}(s) - \min_{s \in \states} \vec{w}(s).
\end{equation*}
The stopping criterion then is given by the condition
\begin{equation} \label{eq:sc_uni}
	\spannorm{\vec{\Delta}_{n}} < \varepsilon. \tag{SC1}
\end{equation}
When the criterion \eqref{eq:sc_uni} is satisfied we have that
\begin{equation} \label{eq:guar_uni}
	\abs{\vec{\Delta}_{n}(s) - \gain} < \varepsilon \qquad\forall s \in \states.
\end{equation}
Moreover, we know that for communicating aperiodic MDPs the criterion \eqref{eq:sc_uni} is satisfied after finitely many steps of Algorithm~\ref{alg:vi}~\cite[Thm.~8.5.2]{Puterman}.
Furthermore, periodic MDPs can be transformed into aperiodic without affecting the average reward.
The transformation works by introducing a self-loop on each state and adapting the rewards accordingly~\cite[Sect.~8.5.4]{Puterman}.
Although this transformation may slow down VI, convergence can now be guaranteed and we can obtain $\varepsilon$-optimal values for any communicating MDP.

The intuition behind this stopping criterion can be explained as follows.
When the computed span norm is small, $\vec{\Delta}_n$ contains nearly the same value in each component.
This means that the difference between the expected ($n-1$)-step and $n$-step total reward is roughly the same in each state.
Since in each state the $n$-step total reward is greedily optimized, there is no possibility of getting more than this difference per step.

Unfortunately, this stopping criterion cannot be applied on general MDPs, as it relies on the fact that all states have the same value, which is not true in general.
Consider for example the MDP of Figure~\ref{fig:example}.
There, we have that $\gain(s_5) = \gain(s_6) = 10$ but $\gain(s_3) = \gain(s_4) = 5$.

In~\cite[Sect.~9.4.2]{Puterman}, it is conjectured that the following criterion may be applicable to general MDPs:
\begin{equation} \label{eq:sc_con1}
	\spannorm{\vec{\Delta}_{n-1}} - \spannorm{\vec{\Delta}_{n}} < \varepsilon. \tag{SC2}
\end{equation}
This stopping criterion requires that the difference of spans becomes small enough.
While investigating the problem, we also conjectured a slight variation:
\begin{equation} \label{eq:sc_con2}
	\norm{\vec{\Delta}_{n} - \vec{\Delta}_{n-1}}_{\infty} < \varepsilon, \tag{SC3}
\end{equation}
where $\norm{\vec{w}}_\infty = \max_{s\in \states} \vec{w}(s)$.
Intuitively, both of these criteria try to extend the intuition of the communicating criterion to general MDPs, i.e.\ to require that in each state the reward gained per step stabilizes.
Example~\ref{ex:break} however demonstrates that neither \eqref{eq:sc_con1} nor \eqref{eq:sc_con2} is a valid stopping criterion.
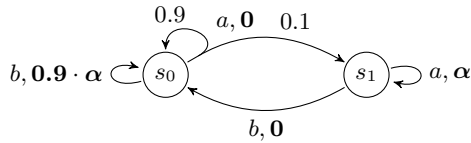
\begin{figure}[t]
	\centering
	\begin{tikzpicture}[auto]
		\node[state] (state_0) {$s_0$};
		\node[state,right=2cm of state_0] (state_1) {$s_1$};
		\coordinate[at= ($(state_0) + (30:0.5)$),label={[label distance=7pt] 60:$a,\mathbf{0}$}] (state_0_a);
		\draw[-] (state_0) -- (state_0_a);
		\path[->]
			(state_0_a) edge[out=30, in=90,looseness=3] node[anchor=south east] {$0.9$} (state_0)
			(state_0_a) edge[out=30, in=150]  node {$0.1$} (state_1)
			(state_0)   edge[loop left]       node {$b, \rewformat{0.9}\cdot\boldsymbol\alpha$}  (state_0)
			(state_1)   edge[loop right]      node {$a, \boldsymbol\alpha$} (state_1)
			(state_1)   edge[out=210, in=-30] node {$b, \rewformat{0}$}  (state_0);
	\end{tikzpicture}
	\caption{A~communicating MDP parametrized by the value $\alpha$.} \label{fig:mc_though}
\end{figure}
\begin{example} \label{ex:break}
	Consider the (aperiodic communicating) MDP in Figure~\ref{fig:mc_though} with a parametrized reward value $\alpha \geq 0$.
	The optimal average reward is $\gain = \alpha$.
	But the first three vectors computed by value iteration are $\vec{t}_0 = (0, 0), \vec{t}_1 = (0.9 \cdot \alpha, \alpha), \vec{t}_2 = (1.8 \cdot \alpha, 2 \cdot \alpha)$.
	Thus, the values of $\vec{\Delta}_1 = \vec{\Delta}_2 = (0.9 \cdot \alpha, \alpha)$ coincide, which means that for every choice of $\varepsilon$ both stopping criteria \eqref{eq:sc_con1} and \eqref{eq:sc_con2} are satisfied by the third iteration.
	However, by increasing the value of $\alpha$ we can make the difference between the average reward $\vec{v}$ and $\vec{\Delta}_2$ arbitrary large, so no guarantee like in Equation~\eqref{eq:guar_uni} is possible.
	\qee
\end{example}

\subsection{Local value iteration} \label{sec:lvi}

In order to remedy the lack of stopping criteria, we provide a modification of VI using MEC decomposition which is able to provide us with an $\varepsilon$-optimal result, utilizing the principle of Equation \eqref{eq:decomp}.
The idea is that for each MEC we compute an $\varepsilon$-optimal value, then consider these values fixed and propagate them through the MDP quotient.

Apart from providing a stopping criterion, this has another practical advantage.
Observe that the naive algorithm updates all states of the model even if the approximation in a single MEC has not $\varepsilon$-converged.
The same happens even when all MECs are already $\varepsilon$-converged and the values only need to propagate along the transient states.
These additional updates of already $\varepsilon$-converged states may come at a high computational cost.
Instead, our method adapts to the potentially very different speeds of convergence in each MEC.

The propagation of the MEC values can be done efficiently by transforming the whole problem to a reachability instance on a modified version of the MEC quotient, which can be solved by, for instance, VI.
We call this variant the \emph{weighted MEC quotient}.
To obtain this weighted quotient, we assume that we have already computed approximate values $w(M)$ of each MEC $M$.
We then collapse the MECs as in the MEC quotient but furthermore introduce new states $s_+$ and $s_-$, which can be reached from each collapsed state by a special action $\mathsf{stay}$ with probabilities corresponding to the approximate value of the MEC.
Intuitively, by taking this action the strategy decides to \enquote{stay} in this MEC and obtain the average reward of the MEC.

Formally, we define the function $f$ as the normalized approximated value, i.e.\ for some MEC $M_i$ we set $f(\hat{s}_i) = \frac{1}{\rmax} w(M_i)$, so that it takes values in $[0, 1]$.
Then, the probability of reaching $s_+$ upon taking the $\mathsf{stay}$ action in $\hat{s}_i$ is defined as $f(\hat{s}_i)$ and dually the transition to $s_-$ is assigned $1 - f(\hat{s}_i)$ probability.
If for example some MEC $M$ had a value $v(M) = \frac{2}{3} \rmax$, we would have that $\trans(\hat{s}, \mathsf{stay}, s_+) = \frac{2}{3}$.
This way, we can interpret reaching $s_+$ as obtaining the maximal possible reward, and reaching $s_-$ to obtaining no reward.
With this intuition, we show in Theorem~\ref{th:quo} that the problem of computing the average reward is reduced to computing the value of each MEC and determining the maximum probability of reaching the state $s_+$ in the weighted MEC quotient.
\begin{definition}[Weighted MEC quotient] \label{def:wmq}
	Let $\widehat\Mdp = (\widehat\states, \hat{s}_\textrm{init}, \widehat\actions, \widehat\av, \widehat\trans, \widehat\rew)$ be the MEC quotient of an MDP $\Mdp$ and let $\mec_{\widehat{S}} = \set{\hat{s}_1, \dots, \hat{s}_n}$ be the set of collapsed states.
	Further, let $f :\mec_{\widehat{S}} \to [0,1]$ be a function assigning a value to every collapsed state.
	We define the \emph{weighted MEC quotient of $\Mdp$ and $f$} as the MDP $\Mdp^f = (\states^f, \initstate^f, \widehat\actions \union \set{\mathsf{stay}}, \av^f, \trans^f, \rew^f)$, where
	\begin{itemize}
		\item $\states^f = \widehat\states \union \set{s_+, s_-}$,
		\item $\initstate^f = \hat{s}_\textrm{init}$,
		\item $\av^f$ is defined as
		\begin{align*}
			\forall \hat{s} \in \widehat\states.~& \av^f(\hat{s}) = \begin{dcases*}
				\widehat\av(\hat{s}) \union \set{\mathsf{stay}} &if $\hat{s} \in \mec_{\widehat{S}},$ \\
				\widehat\av(\hat{s}) &otherwise,
			\end{dcases*} \\
			& \av^f(s_+) = \av^f(s_-) = \emptyset,
		\end{align*}
		\item $\trans^f$ is defined as
		\begin{align*}
			\forall \hat{s} \in \widehat\states, \hat{a} \in \widehat\actions \setminus \set{\mathsf{stay}}.~& \trans^f(\hat{s}, \hat{a}) = \widehat\trans(\hat{s}, \hat{a}) \\
			\forall \hat{s}_i \in \mec_{\widehat{S}}.~& \trans^f(\hat{s}_i, \mathsf{stay}) = \set{s_+ \mapsto f(\hat{s}_i, s_- \mapsto 1 - f(\hat{s}_i)},
		\end{align*}
		\item and the reward function $\rew^f(\hat{s}, \hat{a})$ is chosen arbitrarily (e.g.\ $0$ everywhere), since we only consider a reachability problem on $\Mdp^f$.
	\end{itemize}
\end{definition}
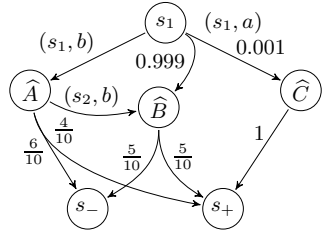
\begin{figure}[t]
	\centering
	\scalebox{0.9}{
		\begin{tikzpicture}[align=center]
			\node[state] (s0) at (-3,0.5) {$s_1$};
			\node[state] (s2) at (-1,-0.5) {$\widehat C$};
			\node[state] (s3) at (-3.1,-0.75) {$\widehat B$};
			\node[state] (s5) at (-5,-0.5) {$\widehat A$};
			\node[state] (sp) at (-2.15,-2.25) {$s_+$};
			\node[state] (sm) at (-4.15,-2.25) {$s_-$};
			
			\node[] at (-2,0.5) {$(s_1,a)$};
			
			\path[->]
			(s0) edge node[anchor=south west,inner sep=1pt] {$0.001$} (s2)
			(s0) edge[out=-30,in=45] node[left]{$0.999$} (s3)
			(s0) edge node[above left, inner sep=1pt] {$(s_1, b)$} (s5)
			(s5) edge[out=-30,in=190] node[above] {$(s_2, b)$} (s3)
			(s5) edge[out=-80,in=170,looseness=0.75] node[anchor=south, inner sep=5pt, pos=0.3]{$\frac{4}{10}$} (sp)
			(s5) edge[out=-80,in=120,looseness=0.5] node[left]{$\frac{6}{10}$} (sm)
			(s2) edge node[left,pos=0.3] {$1$} (sp)
			(s3) edge[out=-90,in=150] node[anchor=south west,inner sep=0pt]{$\frac{5}{10}$} (sp)
			(s3) edge[out=-90,in=30]  node[anchor=south east,inner sep=0pt]{$\frac{5}{10}$} (sm);
		\end{tikzpicture}
	}
	\caption{The weighted quotient of the MDP in Figure~\ref{fig:example} and function $f = \set*{\widehat{A} \mapsto \frac{4}{10}, \widehat{B} \mapsto \frac{5}{10}, \widehat{C} \mapsto \frac{10}{10}}$.
	Rewards and $\mathsf{stay}$ action labels omitted for readability.}
	\label{fig:rquo}
\end{figure}
\begin{example}
	Consider the MDP in Figure~\ref{fig:example}.
	The average rewards of the MECs are $\gain = \set*{\widehat{A} \mapsto 4, \widehat{B} \mapsto 5, \widehat{C} \mapsto 10}$.
	With $f$ defined as in Theorem~\ref{th:quo}, Figure~\ref{fig:rquo} shows the weighted MEC quotient $\Mdp^f$.
	\qee
\end{example}
\begin{theorem} \label{th:quo}
	Given an MDP $\Mdp$ with MECs $\mec(\Mdp) = \set*{M_1, \dots, M_n}$, define $f(\hat{s}_i) = \tfrac{1}{\rmax} \gain(M_i)$ the function mapping each MEC $M_i$ to its value.
	Moreover, let $\Mdp^f$ be the weighted MEC quotient of $\Mdp$ and $f$.
	Then
	\begin{equation*}
		\gain(\initstate) = \rmax \cdot \sup_{\straa \in \straas} \pr^\straa_{\Mdp^f, \initstate^f}(\Diamond s_+).
	\end{equation*}
\end{theorem}
\begin{algorithm}[t]
	\caption{\textsc{LocalVI}} \label{alg:lvi}
	\begin{algorithmic}[1]
		\Require MDP $\Mdp = (\states,\initstate,\actions,\av,\trans,\rew)$, precision $\varepsilon > 0$
		\Ensure $w$, s.t.\ $\abs{w - \gain(\initstate)} < \varepsilon$
		\State $f = \emptyset$
		\For {$M_i = (T_i, A_i) \in \mec(\Mdp)$} \Comment Determine values for MECs
		\State Compute the average reward $w(M_i)$ on $M$, such that $\abs*{w(M_i) - \gain(M_i)} < \tfrac{1}{2} \varepsilon$,
		\State $f(\hat{s}_i) \gets \frac{1}{\rmax} w(M_i)$
		\EndFor
		\State $\Mdp^f\gets$ the weighted MEC quotient of $\Mdp$ and $f$
		\State Compute $p$ s.t.\ $\abs*{p - \sup_{\straa \in \straas} \pr^\straa_{\Mdp^f, \initstate^f}(\Diamond s_+)} < \frac{1}{2 \rmax} \varepsilon$ \Comment Determine reachability
		\State \Return $\rmax \cdot p$
	\end{algorithmic}
\end{algorithm}
The corresponding algorithm is shown in Algorithm~\ref{alg:lvi}.
It takes an MDP and the required precision $\varepsilon$ as input and returns a value $w$, which is $\varepsilon$-close to the average reward $\gain(\initstate)$.
In the first part, for each MEC $M$ the algorithm computes an approximate average reward $w(M)$ and assigns it to the function $f$ (normalized by $\rmax$).
Every MEC is a communicating MDP, therefore the value $w(M)$ can be computed using the naive VI with \eqref{eq:sc_uni} as the stopping criterion.
In the second part, the weighted MEC quotient of $\Mdp$ and $f$ is constructed and the maximum probability $p$ of reaching $s_+$ in $\Mdp^f$ is approximated.
\begin{theorem} \label{th:lvi_corr}
	For every MDP $\Mdp$ and $\varepsilon > 0$, Algorithm~\ref{alg:lvi} terminates and is correct, i.e.\ returns a value $w$, s.t.\ $\abs{w - \gain(\initstate)} < \varepsilon$.
\end{theorem}
For the correctness, we require that $p$ is $\frac{\varepsilon}{2\rmax}$-close to the real maximum probability of reaching $s_+$.
This can be achieved by using the VI algorithms for reachability from \cite{atva} or \cite{haddad2014reachability}, which guarantee error bounds on the computed probability.
Note that $p$ can also be computed by other methods, such as linear programming.
In Section~\ref{sec:exper} we empirically compare these approaches.
%
%

\subsection{On-demand value iteration} \label{sec:odv}

Observe that in Algorithm~\ref{alg:lvi}, the approximations for all MECs are equally precise, irrespective of the effect a MEC's value has on the overall value of the MDP.
Moreover, the whole model is stored in memory and all the MECs are computed beforehand, which can be expensive for large MDPs.
Often this is unnecessary, as we illustrate in the following example.
\begin{example}
	There are three MECs $\widehat{A}, \widehat{B}, \widehat{C}$ in the MDP of Figure~\ref{fig:example}.
	Furthermore, we have that $\pr^{\straa}_{\initstate}(\Diamond \widehat{C}) \leq 0.001$.
	By using the intuition of Equation \eqref{eq:decomp}, we see that no matter where in the interval $[0, \rmax=20]$ its value lies, it contributes to the overall value $\gain(\initstate)$ at most by $0.001 \cdot \rmax = 0.02$.
	If the required precision were $\varepsilon = 0.1$, the effort invested in computing the value of $\widehat{C}$ would not pay off at all and one can completely omit constructing $\widehat{C}$.

	Further, suppose that $\widehat{A}$ was a more complicated MEC, but after a few iterations the criterion \eqref{eq:sc_uni} already shows that the value of $\widehat{A}$ is at most $4.4$.
	Similarly, after several iterations in $\widehat{B}$, we might see that the value of $\widehat{B}$ is greater than $4.5$.
	In this situation, there is no point in further approximating the value of $\widehat{A}$ since the action $b$ leading to it will not be optimal anyway, and its precise value will not be reflected in the result.
	\qee
\end{example}
To eliminate these inefficient updates, we employ the methodology of \emph{bounded real-time dynamic programming} (BRTDP)~\cite{DBLP:conf/icml/McMahanLG05} adapted to the undiscounted setting in~\cite{atva}.
The word \emph{bounded} refers to keeping and updating both a lower and an upper bound on the final result.
It has been shown in \cite{Puterman,ChatterjeeI14} that bounds for the value of a MEC can be derived from the current maximum and minimum of the approximations of VI.
The idea of the BRTDP approach is to perform updates not repetitively for all states in a fixed order, but more often on the \emph{more important} states.
Technically, finite runs of the system are sampled, and updates to the bounds are propagated only along the states of the current run.
Since successors are sampled according to the transition probabilities, the frequently visited (and thus updated) states are those with high probability of being reached, and therefore also having more impact on the result.
In order to guarantee convergence, the non-determinism is resolved by taking the \emph{most promising action}, i.e.\ the one with the current highest upper bound.
Intuitively, when after subsequent updates such an action turns out to be worse than hoped for, its upper bound decreases and a more promising action is chosen next time.

Since BRTDP of~\cite{atva} is developed only for MDP with the reachability (and LTL) objective, we decompose our problem into a reachability and MEC analysis part.
In order to avoid pre-computation of all MECs with the same precision, we instead compute the values for each MEC only when they could influence the long-run average reward starting from the initial state.
Intuitively, the more a particular MEC is encountered while sampling, the more it is \enquote{reached} and the more precise information we require about its value.

To achieve this, we store upper and lower bounds on its value in the functions $u$ and $\lowerbound$ and refine them on demand by applying VI.
We modify the definition of the weighted MEC quotient to incorporate these lower and upper bounds by introducing the state $s_?$ (in addition to $s_+, s_-$).
We call this construction the \emph{bounded MEC quotient}.
Intuitively, the probability of reaching $s_+$ from a collapsed state now represents the lower bound on its value, while the probability of reaching $s_?$ describes the gap between the upper and lower bound.


\begin{definition}[Bounded MEC quotient] \label{def:bquo}
	Let $\widehat\Mdp = (\widehat\states, \hat{s}_\textrm{init}, \widehat\actions, \widehat\av, \widehat\trans, \widehat\rew)$ be the MEC quotient of an MDP $\Mdp$ with collapsed states $\mec_{\widehat{\states}} = \set{\hat{s}_1, \dots, \hat{s}_n}$ and let $\lu: \set{\hat{s}_1, \dots, \hat{s}_n} \to [0,1]$ be functions that assign a lower and upper bound, respectively, to every collapsed state in $\widehat\Mdp$.
	The \emph{bounded MEC quotient $\Mdp^{\lu}$ of $\Mdp$ and $\lu$} is defined as in Definition~\ref{def:wmq} with the following changes.
	\begin{itemize}
		\item $\states^{\lu} = \widehat\states \union \set{s_?}$,
		\item $\av^{\lu}(s_?) = \emptyset$,
		\item $\forall \hat{s} \in \mec_{\widehat{S}}.~\trans^{\lu}(\hat{s}, \mathsf{stay}) = \set*{s_+ \mapsto \lowerbound(\hat{s}), s_- \mapsto 1 - \upperbound(\hat{s}), s_? \mapsto \upperbound(\hat{s}) - \lowerbound(\hat{s})}$.
	\end{itemize}
	The unshortened definition can be found in \cite[Appendix~D]{techrep}. 
\end{definition}
The probability of reaching $s_+$ and the probability of reaching $\set*{s_+, s_?}$ give the lower and upper bound on the value $\gain(\initstate)$, respectively.
\begin{corollary} \label{th:quo_bounded}
	Let $\Mdp$ be an MDP and $\lu$ functions mapping each MEC $M_i$ of $\Mdp$ to (normalized) lower and upper bounds on the value, respectively, i.e.\ $\lowerbound(\hat{s}_i) \leq \frac{1}{\rmax} \gain(M_i) \leq \upperbound(\hat{s}_i)$.
	Then
	\begin{equation*}
		\rmax \cdot \sup_{\straa \in \straas} \pr^\straa_{\Mdp^{\lu}, \initstate^{\lu}}(\Diamond s_+) \leq v(\initstate) \leq \rmax \cdot \sup_{\straa \in \straas} \pr^\straa_{\Mdp^{\lu}, \initstate^{\lu}}(\Diamond \set*{s_+, s_?}),
	\end{equation*}
	where $\Mdp^{\lu}$ is the bounded MEC quotient of $\Mdp$ and $\lu$.
\end{corollary}
%
\begin{algorithm}[t]
	\caption{\textsc{OnDemandVI}}\label{alg:skeleton}
	\begin{algorithmic}[1]
		\Require MDP $\Mdp = (\states, \initstate, \actions, \av, \trans, \rew)$, precision $\varepsilon > 0$, threshold $k \geq 2$
		\Ensure $w$, s.t.\ $\abs{w - v(\initstate)} < \varepsilon$
		\State Set $\upperbound(\cdot, \cdot) \gets 1$, $\upperbound(s_-, \cdot) \gets 0$; $\lowerbound(\cdot, \cdot) \gets 0$, $\lowerbound(s_+, \cdot) \gets 1$ \Comment Initialize
		\State Let $A(s) := \argmax_{a \in \av^{\lu}(s)} \upperbound(s, a)$
		\State Let $\upperbound(s) := \max_{a \in A(s)} \upperbound(s, a)$ and $\lowerbound(s) := \max_{a \in A(s)} \lowerbound(s, a)$
		\Repeat
			\State $s \gets \initstate^{\lu}, \fpath \gets s$ \Comment Generate path
			\Repeat \label{ln:explore-begin}
				\State $a \gets$ sampled uniformly from $A(s)$
				\State $s \gets$ sampled according to $\Delta^{\lu}(s,a)$
				\State $\fpath \gets \fpath, a, s$
			\Until $s \in \set{s_+,s_-,s_?}$ or $\appear(s, \fpath) = k$ \Comment Terminate path \label{ln:explore-end}
			
			\If {$\pop(\fpath) = s_?$} \label{ln:refine-begin} \Comment Refine MEC in which $\mathsf{stay}$ was taken
				\State $\pop(\fpath)$
				\State $\widehat{q} \gets$ $\topx(\fpath)$
				\State Run VI on $\widehat{q}$, updating $\upperbound$ and $\lowerbound$, until $\upperbound - \lowerbound$ is halved
				\State Update $\trans^{\lu}(\widehat{q}, \mathsf{stay})$ according to Definition~\ref{def:bquo}
				\label{ln:refine-end}
			\ElsIf {$\appear(s, \fpath) = k$} \Comment Update EC-collapsing
				\State \textsc{OnTheFlyEc} \label{alg:line:collapse}
			\EndIf
			\Repeat \label{ln:backpropagate-begin} \Comment Back-propagate values
				\State $a \gets \pop(\fpath)$, $s \gets \pop(\fpath)$
				\State $\upperbound(s, a) \gets \sum_{s' \in S} \Delta(s, a, s') \cdot \upperbound(s')$
				\State $\lowerbound(s, a) \gets \sum_{s' \in S} \Delta(s, a, s') \cdot \lowerbound(s')$
			\Until $\fpath = \emptyset$ \label{ln:backpropagate-end}
		\Until $\upperbound(\initstate) - \lowerbound(\initstate) < \frac{2 \varepsilon}{\rmax}$ \Comment {Terminate}
		\State \Return $\rmax \cdot \tfrac{1}{2} (\upperbound(\initstate) + \lowerbound(\initstate))$
	\end{algorithmic}
\end{algorithm}
\begin{algorithm}[t]
	\makeatletter \renewcommand{\ALG@name}{Procedure} \makeatother
	\caption{\textsc{OnTheFlyEc}}\label{alg:fac}
	\begin{algorithmic}[1]
		\For {$(T_i,A_i) \in \mec(\Mdp^{\lu})$}
			\State Collapse $(T_i,A_i)$ to $\hat{s}_i$ in $\Mdp^{\lu}$
			\For {$s\in T_i, a\in \av(s)\setminus A_i$}
				\State $\upperbound(\hat{s}_i, (s,a)) \gets \upperbound(s,a)$
				\State $\lowerbound(\hat{s}_i, (s,a)) \gets \lowerbound(s,a)$
			\EndFor
			\State Add the $\mathsf{stay}$ action according to Definition~\ref{def:bquo}.
		\EndFor
	\end{algorithmic}
\end{algorithm}
Algorithm~\ref{alg:skeleton} shows the on-demand VI.
The implementation maintains a partial model of the MDP and $\Mdp^{\lu}$, which contains only the states explored by the runs.
It interleaves two concepts: (i)~naive VI is used to provide upper and lower bounds on the value of discovered end components, (ii)~the method of~\cite{atva} is used to compute the reachability on the collapsed MDP.

In lines~\ref{ln:explore-begin}--\ref{ln:explore-end} a random run is sampled following the \enquote{most promising} actions, i.e.\ the ones with maximal upper bound.
The run terminates once it reaches $s_+, s_-$ or $s_?$, which only happens if $\mathsf{stay}$ was one of the most promising actions.
A~likely arrival to $s_?$ reflects a high difference between the upper and lower bound and, if the run ends up in $s_?$, this indicates that the upper and lower bounds of the MEC probably have to be refined.
Therefore, in lines~\ref{ln:refine-begin}--\ref{ln:refine-end} the algorithm resumes VI on the corresponding MEC to get a more precise result.
This decreases the gap between the upper and lower bound for the corresponding collapsed state, thus decreasing the probability of reaching $s_?$ again.

The algorithm uses the function $\appear(s, \fpath) = \abs{\set{i \in \Naturals \mid s = \fpath[i]}}$ to count the number of occurrences of the state $s$ on the path $\fpath$.
Whenever we encounter the same state $k$ times (where $k$ is given as a parameter), this indicates that the run may have got stuck in an end component.
In such a case, the algorithm calls \textsc{OnTheFlyEc}~\cite{atva}, presented in Procedure~\ref{alg:fac}, to detect and collapse end components of the partial model.
By calling \textsc{OnTheFlyEc} we compute the bounded quotient of the MDP on the fly.
Without collapsing the end components, our reachability method could remain forever in an end component, and thus never reach $s_+$, $s_-$ or $s_?$.
Finally, in lines~\ref{ln:backpropagate-begin}--\ref{ln:backpropagate-end} we back-propagate the upper and lower bounds along the states of the simulation run.
%
\begin{theorem} \label{th:skeleton_corr}
	For every MDP $\Mdp$, $\varepsilon > 0$ and $k \geq 2$, Algorithm~\ref{alg:skeleton} terminates almost surely and is correct, i.e.\ returns a value $w$, s.t.\ $\abs{w - v(\initstate)} < \varepsilon$.
\end{theorem}
\section{Implementation and experimental results} \label{sec:exper}

In this section, we compare the runtime of our presented approaches to established tools.
All benchmarks have been run on a 4.4.3-gentoo x64 virtual machine with 3.0 GHz per core, a time limit of one hour and memory limit of 8GB.
The precision requirement for all approximative methods is $\varepsilon = 10^{-6}$.
We implemented our constructions as a package in the PRISM Model Checker \cite{KNP11}.
We used the 64-bit Oracle JDK version \texttt{1.8.0\_102-b14} as Java runtime for all executions.
All measurements are given in seconds, measuring the total user CPU time of the PRISM process using the UNIX tool \texttt{time}.

\subsection{Models}

First, we briefly explain the examples used for evaluation.
\textbf{virus} \cite{kwiatkowska2009probabilistic} models a virus spreading through a network.
We reward each attack carried out by an infected machine.
Note that in this model, no machine can \enquote{purge} the virus, hence eventually all machines will be infected.
\textbf{cs\_nfail} \cite{komuravelli2012assume} models a client-server mutual exclusion protocol with probabilistic failures of the clients.
A~reward is given for each successfully handled connection.
\textbf{investor} \cite{MM07,MM02} models an investor operating in a stock market.
The investor can decide to sell his stocks and keep their value as a reward or hold them and wait to see how the market evolves.
The rewards correspond to the value of the stocks when the investor decides to sell them, so maximizing the average reward corresponds to maximizing the expected selling value of the stocks.
\textbf{phil\_nofair} \cite{DFP04} represents the (randomised) dining philosophers without fairness assumptions.
We use two reward structures, one where a reward is granted each time a philosopher \enquote{thinks} or \enquote{eats}, respectively.
\textbf{rabin} \cite{rabin1982n} is a well-known mutual exclusion protocol, where multiple processes repeatedly try to access a shared critical section.
Each time a process successfully enters the critical section, a reward is given.
\textbf{zeroconf} \cite{KNPS06} is a network protocol designed to assign IP addresses to clients without the need of a central server while still avoiding address conflicts.
We explain the reward assignment in the corresponding result section.
\textbf{sensor} \cite{komuravelli2012assume} models a network of sensors sending values to a central processor over a lossy connection.
A reward is granted for every \emph{work} transition.

\subsection{Tools}

We will compare several different variants of our implementations, which are described in the following.
\begin{itemize}
	\item Naive value iteration (\texttt{NVI}) runs the value iteration on the whole MDP as in Algorithm~\ref{alg:vi} of Section~\ref{sec:nvi} together with the stopping criterion \eqref{eq:sc_con1} conjectured by \cite[Sect.~9.4.2]{Puterman}.
	As the stopping criterion is incorrect, we will not only include the runtime until the stopping criterion is fulfilled, but also until the computed value is $\varepsilon$-close to the known solution.
	\item Our MEC decomposition approach presented in Algorithm~\ref{alg:lvi} of Section~\ref{sec:lvi} is denoted by \texttt{MEC-\textit{reach}}, where $\texttt{\textit{reach}}$ identifies one of the following reachability solver used on the quotient MDP.
	\begin{itemize}
		\item PRISM's value iteration (\texttt{VI}), which iterates until none of the values change by more than $10^{-8}$.
		While this method is theoretically imprecise, we did not observe this behaviour in our examples.\footnote{PRISM contains several other methods to solve reachability, which all are imprecise and behaved comparably in our tests.}
		\item An exact reachability solver based on linear programming (\texttt{LP})~\cite{Giro14}.
		\item The BRTDP solver with guaranteed precision of~\cite{atva} (\texttt{BRTDP}).
		This solver is highly configurable.
		Among others, one can specify the heuristic which is used to resolve probabilistic transitions in the simulation.
		This can happen according to transition probability (\texttt{PR}), round-robin (\texttt{RR}) or maximal difference (\texttt{MD}).
		Due to space constraints, we only compare to the \texttt{MD} exploration heuristic here.
		Results on the other heuristics can be found in \cite[Appendix~E]{techrep} 
	\end{itemize}
	\item \texttt{ODV} is the implementation of the on-demand value iteration as in Algorithm~\ref{alg:skeleton} of Section~\ref{sec:odv}.
	Analogously to the above, we only provide results on the \texttt{MD} heuristic here.
	The results on \texttt{ODV} together with the other heuristics can also be found in \cite[Appendix~E]{techrep}. 
\end{itemize}
Furthermore, we will compare our methods to the state-of-the-art tool MultiGain, version 1.0.2~\cite{DBLP:conf/tacas/BrazdilCFK15} abbreviated by \texttt{MG}.
MultiGain uses linear programming to exactly solve mean payoff objectives among others.
We use the commercial LP solver Gurobi 7.0.1 as backend\footnote{MultiGain also supports usage of the LP solver \texttt{lp\_solve 5.5} bundled with PRISM, which consistently performed worse than the Gurobi backend.}.
We also instantiated \texttt{\textit{reach}} by an implementation of the interval iteration algorithm presented in~\cite{haddad2014reachability}.
This variant performed comparable to \texttt{MEC-VI} and therefore we omitted it.

\subsection{Results}

\begin{table}[t]
	\centering
	\caption{Runtime comparison of our approaches to MultiGain on various, reasonably sized models.
		Timeouts (1h) are denoted by TO.
		Strongly connected models are denoted by \enquote{scon} in the MEC column.
		The best result in each row is marked in bold, excluding NVI due to its imprecisions.
		For NVI, we list both the time until the stopping criterion is satisfied and until the values actually converged.}
	\setlength{\tabcolsep}{-1pt}
	\begin{tabu} to \linewidth {lRRCCCCcC}
		\textbf{Model} & \textbf{States} & \textbf{MECs} & \texttt{MG} & \texttt{NVI} & \texttt{MEC-VI} & \texttt{MEC-LP} & \texttt{MEC-BRTDP} & \texttt{ODV} \\
		\toprule
		virus        &    809 &       1 &  \textbf{3.76} & 3.50/3.71 & 4.09 &         4.41 &         4.40 &   TO \\
		cs\_nfail4   &    960 &     176 &  4.86 &   10.2/TO & \textbf{4.38} &           TO &         9.39 & 16.0 \\
		investor     &   6688 &     837 & 16.75 &   4.23/TO & \textbf{8.83} &           TO &         64.5 & 18.7 \\
		phil-nofair5 &  93068 &    scon &    TO & 23.5/30.3 &   \textbf{70} &  \textbf{70} &  \textbf{70} &   TO \\
		rabin4       & 668836 &    scon &    TO &  87.8/164 &  \textbf{820} & \textbf{820} & \textbf{820} &   TO \\
		\bottomrule
	\end{tabu}
	\label{tbl:experiment_mg}
\end{table}

The experiments outlined in Table~\ref{tbl:experiment_mg} show that our methods outperform MultiGain significantly on most of the tested models.
Furthermore, we want to highlight the \textbf{investor} model to demonstrate the advantage of \texttt{MEC-VI} over \texttt{MEC-LP}.
With higher number of MECs in the initial MDP, which is linked to the size of the reachability LP, the runtime of \texttt{MEC-LP} tends to increase drastically, while \texttt{MEC-VI} performs quite well.
Additionally, we see that \texttt{NVI} fails to obtain correct results on any of these examples.

\texttt{ODV} does not perform too well in these tests, which is primarily due to the significant overhead incurred by building the partial model dynamically.
This is especially noticeable for strongly connected models like \textbf{phil-nofair} and \textbf{rabin}.
For these models, every state has to be explored and \texttt{ODV} does a lot of superfluous computations until the model has been explored fully.
On \textbf{virus}, the bad performance is due to the special topology of the model, which obstructs the back-propagation of values.

Moreover, on the two strongly connected models all MEC decomposition based methods perform worse than naive value iteration as they have to obtain the MEC decomposition first.
Furthermore, all three of those methods need the same amount of for these models, as the weighted MEC quotient only has a single state (and the two special states), thus the reachability query is trivial.

\begin{table}[t]
	\centering
	\caption{Runtime comparison of our on-demand VI method with the previous approaches.
		All of those behaved comparable to \texttt{MEC-VI} or worse, and due to space constraints we omit them.
		MO denotes a memory-out.
		Aside from runtime, we furthermore list the number of explored states and MECs of \texttt{ODV}}
	\begin{tabu} to \linewidth {lRCCCC}
		\textbf{Model}      & \textbf{States} & \texttt{MEC-VI} & \texttt{ODV} & \texttt{ODV} States & \texttt{ODV} MECs \\
		\toprule
		zeroconf(40,10)     &         3001911 &              MO &         5.05 &                 481 &                 3 \\
		\quad\emph{avoid}   &                 &                 &              &                 582 &                 3 \\
		zeroconf(300,15)    &         4730203 &              MO &         16.6 &                 873 &                 3 \\
		\quad\emph{avoid}   &                 &                 &              &                5434 &                 3 \\
		sensors(2)          &            7860 &            18.9 &         20.1 &                3281 &               917 \\
		sensors(3)          &           77766 &            2293 &         37.2 &               10941 &              2301 \\
		\bottomrule
	\end{tabu}
	\label{tbl:experiment_big}
\end{table}

In Table~\ref{tbl:experiment_big} we present results of some of our methods on \textbf{zeroconf} and \textbf{sensors}, which both have a structure better suited towards \texttt{ODV}.
The \textbf{zeroconf} model consists of a big transient part and a lot of \enquote{final} states, i.e. states which only have a single self-loop.
\textbf{sensors} contains a lot of small, often unlikely-to-be-reached MECs.

On the \textbf{zeroconf} model, we evaluate the average reward problem with two reward structures.
In the default case, we assign a reward of 1 to every final state and zero elsewhere.
This effectively is solving the reachability question and thus it is not surprising that our method gives similarly good results as the BRTDP solver of~\cite{atva}.
The \emph{avoid} evaluation has the reward values flipped, i.e. all states except the final ones yield a payoff of 1.
With this reward assignment, the algorithm performed slightly slower, but still extremely fast given the size of the model.
We also tried assigning pseudo-random rewards to every non-final state, which did not influence the speed of convergence noticeably.
We want to highlight that the mem-out of \texttt{MEC-VI} already occurred during the MEC-decomposition phase.
Hence, no variant of our decomposition approach can solve this problem.

Interestingly, the naive value iteration actually converges on \textbf{zeroconf}(40,10) in roughly 20 minutes.
Unfortunately, as in the previous experiments, the used incorrect stopping criterion was met a long time before that.

Further, when comparing \textbf{sensors}(2) to \textbf{sensors}(3), the runtime of \texttt{ODV} only doubled, while the number of states in the model increased by an order of magnitude and the runtime of \texttt{MEC-VI} even increased by two orders of magnitude.

These results show that for some models, \texttt{ODV} is able to obtain an $\varepsilon$-optimal estimate of the mean payoff while only exploring a tiny fraction of the state space.
This allows us to solve many problems which previously were intractable simply due to an enormous state space.

\section{Conclusion}

We have discussed the use of value iteration for computing long-run average rewards in general MDPs.
We have shown that the conjectured stopping criterion from literature is not valid, designed two modified versions of the algorithm and have shown guarantees on their results.
The first one relies on decomposition into VI for long-run average on separate MECs and VI for reachability on the resulting quotient, achieving global error bounds from the two local stopping criteria.
The second one additionally is simulation-guided in the BRTDP style, and is an anytime algorithm with a stopping criterion.
The benchmarks show that depending on the topology, one or the other may be more efficient, and both outperform the existing linear programming on all larger models.
For future work, we pose the question of how to automatically fine-tune the parameters of the algorithms to get the best performance.
For instance, the precision increase in each further call of VI on a MEC could be driven by the current values of VI on the quotient, instead of just halving them.
This may reduce the number of unnecessary updates while still achieving an increase in precision useful for the global result.
\bibliographystyle{alpha}
\bibliography{main}

\appendix
\section*{Appendix}

\section{Proof of Theorem~\ref{th:quo}}
\label{proof:quo}

We use $\Next S$ to denote the event of reaching some state in $S$ in the next step.
\begin{align*}
	\gain(\initstate) & = \sup_{\straa\in \straas} \sum_{M_i \in \mec(\Mdp)} \pr^\straa_{\Mdp, \initstate} (\Diamond\alws M_i) \cdot \gain(M_i) \\
		& = \rmax \cdot \sup_{\straa\in \straas} \sum_{M_i \in \mec(\Mdp)} \pr^\straa_{\Mdp, \initstate} (\Diamond\alws M_i) \cdot f(\hat{s}_i) \\
		& \qquad \text{(since $\gain(M_i)=\rmax\cdot f(\hat{s}_i)$ by assumption)} \\
		& = \rmax \cdot \sup_{\straa\in \straas} \sum_{M_i \in \mec(\Mdp)} \pr^\straa_{\Mdp^f, \initstate^f} \left(\Diamond \left(\hat{s}_i \land \Next \left(s_+ \vee s_- \right) \right) \right) \cdot f(\hat{s}_i) \\
		& = \rmax \cdot\sup_{\straa\in \straas} \sum_{M_i\ \in \mec(\Mdp)} \pr^\straa_{\Mdp^f, \initstate^f} \left(\Diamond \hat{s}_i \right) \cdot \pr^\straa_{\Mdp^f, \hat{s}_i} \left(\Next (s_+ \vee s_-)\right) \cdot f(\hat{s}_i) \\
		& \qquad \text{(by the Markov property)} \\
		& = \rmax \cdot\sup_{\straa\in \straas} \sum_{M_i \in \mec(\Mdp)} \pr^\straa_{\Mdp^f, \initstate^f} (\Diamond \hat{s}_i) \cdot \pr^\straa_{\Mdp^f, \hat{s}_i} (\Next s_+) \\
		& \qquad \text{(since $\pr^\straa_{\Mdp^f, \hat{s}_i}(\Next s_+) = \pr^\straa_{\Mdp^f, \hat{s}_i} (\Next (s_+ \vee s_-)) \cdot f(\hat{s}_i)$)} \\
		& = \rmax \cdot \sup_{\straa\in \straas} \pr^\straa_{\Mdp^f}(\Diamond s_+). \tag*{\mbox{\qed}}
\end{align*}

\section{Proof of Theorem~\ref{th:lvi_corr}}

Algorithm~\ref{alg:lvi} terminates since every part only takes finitely many steps. We now prove correctness, i.e.\ that upon termination the algorithm returns value $p$, such that $\abs{\rmax \cdot p - \gain(\initstate)} < \varepsilon$.
\begin{align*}
	\sup_{\straa \in \straas} \pr^\straa_{\Mdp^f, \initstate^f}(\Diamond s_+)
		& = \sup_{\straa \in \straas} \sum_{M_i\in \mec(\Mdp)} \pr^\straa_{\Mdp, \initstate} (\Diamond \alws M_i) \cdot f(\hat{s}_i) \\
		& \qquad \text{(from the proof of Theorem~\ref{th:quo})} \\
		& \leq \sup_{\straa \in \straas} \sum_{M_i\in \mec(\Mdp)} \pr^\straa_{\Mdp} (\Diamond \alws M_i) \frac{\gain(M_i) + \tfrac{1}{2} \varepsilon}{\rmax} \quad \text{(by assumption)}\\
		& = \frac{1}{\rmax} \cdot \sup_{\straa \in \straas} \sum_{M_i\in \mec(\Mdp)} \pr^\straa_{\Mdp, \initstate} (\Diamond \alws M_i) \cdot \left(\gain(M_i) + {\tfrac{1}{2}} \varepsilon \right)\\
		& \leq \frac{1}{\rmax} \left( \tfrac{1}{2} \varepsilon + \sup_{\straa \in \straas} \sum_{M_i\in \mec(\Mdp)} \pr^\straa_{\Mdp, \initstate}(\Diamond \alws M_i) \cdot \gain(M_i) \right) \\
		& = \frac{\gain(\initstate)}{\rmax} + \frac{\varepsilon}{2\rmax}.
\end{align*}
Analogously we can prove that $\sup_{\straa \in \straas} \pr^\straa_{\Mdp^f, \initstate^f}(\Diamond s_+) \geq  \frac{\gain(\initstate)}{\rmax} - \frac{\varepsilon}{2\rmax}$. Together
\begin{equation}
\label{eq:sth1}
	\abs{\sup_{\straa \in \straas} \pr^\straa_{\Mdp^f, \initstate^f}(\Diamond s_+) - \frac{\gain(\initstate)}{\rmax}} \leq \frac{\varepsilon}{2 \rmax}
\end{equation}
By assumption it also holds that
\begin{equation}
	\abs{p - \sup_{\straa \in \straas} \pr^\straa_{\Mdp^f, \initstate^f}(\Diamond s_+)} < \frac{\varepsilon}{2\rmax} \label{eq:sth2}
\end{equation}
By the triangle inequality we obtain from \eqref{eq:sth1} and \eqref{eq:sth2} that
\begin{equation*}
	\abs{p - \frac{\gain(\initstate)}{\rmax}} < \frac{\varepsilon}{\rmax}.
\end{equation*}
and thus
\begin{equation*}
	\abs{\rmax \cdot p - \gain(\initstate)} < \varepsilon. \tag*{\mbox{\qed}}
\end{equation*}

\section{Proof of Theorem~\ref{th:skeleton_corr}}

We extend the proof of \cite[Thm.~3]{atva}, which can be found in the technical report \cite{DBLP:journals/corr/BrazdilCCFKKPU14}.
To this end, we summarize some properties of $\upperbound$ and $\lowerbound$.
Let $M$ be some MEC and $\hat{s}_i$ its corresponding collapsed state.
\begin{itemize}
	\item $\upperbound(\hat{s}_i)$ and $\lowerbound(\hat{s}_i)$ are always and upper and lower bound of value of $M$.
	This follows from the error bound on VI within $M$.
	\item Let $n_i$ be the number of times the $\mathsf{stay}$ action is taken in $\hat{s}_i$.
	We have
	\begin{equation*}
		\lim_{n_i \to \infty} \left(\upperbound(\hat{s}_i) - \lowerbound(\hat{s}_i)\right) = 0.
	\end{equation*}
	This follows from convergence of VI on MECs and by the fact that if $\mathsf{stay}$ is taken infinitely often, a round of VI on $M$ happens infinitely often.
\end{itemize}
Consequently, the proof of \cite[Thm.~3]{DBLP:journals/corr/BrazdilCCFKKPU14} applies.
\qed

\section{Definition of bounded MEC quotient} \label{sec:bounded_quotient}
\begin{definition}[Bounded MEC quotient]
	Let $\widehat\Mdp = (\widehat\states, \hat{s}_\textrm{init}, \widehat\actions, \widehat\av, \widehat\trans, \widehat\rew)$ be the MEC quotient of an MDP $\Mdp$ with collapsed states $\mec_{\widehat{S}} = \set{\hat{s}_1, \dots, \hat{s}_n}$.
	Further, let $\lu: \set{\hat{s}_1, \dots, \hat{s}_n} \to [0,1]$ be functions assigning a lower and upper bound, respectively, to every collapsed state.
	We define the \emph{bounded MEC quotient of $\Mdp$ and $f$} as the MDP $\Mdp^{\lu} = (\states^{\lu}, \initstate^{\lu}, \widehat\actions \union \set{\mathsf{stay}}, \av^{\lu}, \trans^{\lu}, \rew^{\lu})$, where
	\begin{itemize}
		\item $\states^{\lu} = \widehat\states \union \set{s_+, s_-, s_?}$,
		\item $\initstate^{\lu} = \hat{s}_\textrm{init}$,
		\item $\av^{\lu}$ is defined as
		\begin{align*}
			\forall \hat{s} \in \widehat\states.~& \av^{\lu}(\hat{s}) = \begin{dcases*}
				\widehat\av(\hat{s}) \union \set{\mathsf{stay}} & if $\hat{s} \in \mec_{\widehat{S}}$, \\
				\widehat\av(\hat{s}) & otherwise,
			\end{dcases*} \\
			& \av^{\lu}(s_+) = \av^{\lu}(s_-) = \av^{{\lu}}(s_?) = \emptyset,
		\end{align*}
		\item $\trans^{\lu}$ is defined as
		\begin{gather*}
			\forall \hat{s} \in \widehat\states, \hat{a} \in \widehat\av(\hat{s}) \setminus \set{\mathsf{stay}}.~ \trans^{\lu}(\hat{s}, \hat{a}) = \widehat\trans(\hat{s}, \hat{a}) \\
			\forall \hat{s} \in \mec_{\widehat{S}}.~ \trans^{\lu}(\hat{s}, \mathsf{stay}) = \set*{s_+ \mapsto \lowerbound(\hat{s}), s_- \mapsto 1 - \upperbound(\hat{s}), s_? \mapsto \upperbound(\hat{s}) - \lowerbound(\hat{s})},
		\end{gather*}
		\item and the reward function $\rew^{\lu}(\hat{s}, \hat{a})$ is chosen arbitrarily (e.g.\ $0$ everywhere), since we only consider a reachability problem on $\Mdp^{\lu}$.
	\end{itemize}
\end{definition}


\section{More experimental data} \label{section:exper_data_appendix}

\begin{table}[h]
	\centering
	\caption{Additional data for Table~\ref{tbl:experiment_mg}.
		The results for \texttt{MEC-BRTDP} and \texttt{ODV} using the different heuristics are listed in the order \texttt{PR}, \texttt{RR}, \texttt{MD}.
		Note that for strongly connected (scon) models, all variants of \texttt{MEC-BRTDP} take the same time, since the reachability solver is not called at all.}
	\begin{tabu} to 0.8\linewidth {lrrCC}
		\textbf{Model} & \textbf{States} & \textbf{MECs} & \texttt{MEC-BRTDP} & \texttt{ODV} \\
		\toprule
		virus        &    809 &       1 &   4.69/3.96/4.40 &     TO/2452/TO \\
		cs\_nfail4   &    960 &     176 &   20.3/8.83/9.39 & 51.3/15.4/16.0 \\
		investor     &   6688 &     837 &   47.1/48.6/64.5 & 17.3/21.3/18.7 \\
		phil-nofair5 &  93068 &    scon &               70 &       TO/TO/TO \\
		rabin4       & 668836 &    scon &              820 &       TO/TO/TO \\
		\bottomrule
	\end{tabu}
	\label{tbl:experiment_mg_app}
\end{table}

\begin{table}[h]
	\centering
	\caption{Additional data for Table~\ref{tbl:experiment_big}.
		We list the results of using \texttt{ODV} with the three different mentioned exploration heuristics.
		Heuristics are listed in the order \texttt{PR}, \texttt{RR}, \texttt{MD}.}
	\begin{tabu} to \linewidth{lrCCC}
		\textbf{Model}            & \textbf{States} &            Time &   Explored States &          Explored MECs \\
		\toprule
		zeroconf(40,10)           &         3001911 &  83.4/9.72/5.05 &      516/1643/481 &                  3/3/3 \\
		~~\emph{avoid}            &                 &                 &      905/2475/582 &                  3/3/3 \\
		zeroconf(300,15)          &         4730203 &   1831/294/16.6 &      731/2291/873 &                  3/3/3 \\
		~~\emph{avoid}            &                 &                 &   1595/10759/5434 &                 3/22/3 \\
		sensors(2)                &            7860 &  31.8/26.1/20.1 &    5308/4915/3281 &          1474/1444/917 \\
		sensors(3)                &           77766 &      305/480/37 & 31431/34239/10941 &        9125/11041/2301 \\
		\bottomrule
	\end{tabu}
	\label{tbl:experiment_big_app}
\end{table}


\end{document}